\documentclass[draft]{amsart}
\usepackage{amscd}
\usepackage{amsxtra}
\usepackage{amssymb, amsmath}
\usepackage{amsthm}%With this package one can choose among three types of
\theoremstyle{plain}
\newtheorem{theorem}{Theorem}[subsection]
\newtheorem{proposition}[theorem]{Proposition}
\newtheorem{lemma}[theorem]{Lemma}

\newtheorem{definition}[theorem]{Definition}
\newtheorem{remark}[theorem]{Remark}

\begin{document}

%spaces
\newcommand{\U}{\ensuremath{\mathfrak{u}_{1}}}

%special bundles
\newcommand{\ad}{\ensuremath { ad(\mathfrak{u}_{1})}}
\newcommand{\Oo}{\ensuremath{\varOmega^{0}(ad(\mathfrak{u}_{1}))}}
\newcommand{\Ok}{\ensuremath {\varOmega^{1}(ad(\mathfrak{u}_{1}))}}

%vector bundles
\newcommand{\s}{\ensuremath{\mathcal{S}}}
\newcommand{\csa }{\ensuremath {\mathcal{S_{\alpha}}}}
\newcommand{\cs}{\ensuremath{\mathcal{S^{c}}}}
\newcommand{\vsa}{\ensuremath{\varGamma(\mathcal{S^{+}_{\alpha}})}}
\newcommand{\la }{\ensuremath{{\mathcal{L}}_{\alpha}}}
\newcommand{\sla }{\ensuremath{{\mathcal{L}}^{1/2}_{\alpha}}}
\newcommand{\pcsa }{\ensuremath {\mathcal{S^{+}_{\alpha}}}}

%connection space
\newcommand{\ca}{\ensuremath{\mathcal{C}_{\alpha}}}
\newcommand{\Aa}{\ensuremath {\mathcal{A}_{\alpha}}}
\newcommand{\Q}{\ensuremath{\mathcal{A}_{\alpha}\times_\mathcal{G_{\alpha}}
\varGamma (S^{+}_{\alpha})}}
\newcommand{\B}{\ensuremath{\mathcal{B}_{\alpha}}}

%gauge group
\newcommand{\G}{\ensuremath{\mathcal{G}_{\alpha}}}
\newcommand{\wG}{\ensuremath{\widehat{\mathcal{G}}_{\alpha}}}

%operators
\newcommand{\h}{\ensuremath{H^{\mathcal{S}\mathcal{W}}_{(A,\phi)}}}
\newcommand{\hh}{\ensuremath{\widehat{H}^{\mathcal{S}\mathcal{W}}_{(A,\phi)}}}

%greek letters
\newcommand{\lda}{\lambda}

%structures
\newcommand{\spinc}{\ensuremath{Spin^{c}_{4}}}
\newcommand{\spin }{\ensuremath{Spin_{4}}}

%fields
\newcommand{\Z }{\ensuremath{\mathbb {Z}}}
\newcommand{\R }{\ensuremath{\mathbb {R}}}
\newcommand{\C }{\ensuremath {\mathbb {C}}}

%math symbols
\newcommand{\iso }{\ensuremath {\thickapprox }}

%abbreviations
\newcommand{\sw}{\ensuremath {\mathcal{S}\mathcal{W}}}
\newcommand{\swa}{\ensuremath {\mathcal{S}\mathcal{W}_{\alpha}}}
\newcommand{\y}{\ensuremath {\mathcal{Y}\mathcal{M}}}
\newcommand{\yp}{\ensuremath {\mathcal{Y}\mathcal{M}^{+}}}
\newcommand{\yn}{\ensuremath {\mathcal{Y}\mathcal{M}^{-}}}
\newcommand{\Cf }{\ensuremath {\mathcal{C}_{\alpha}}}
\newcommand{\w }{\ensuremath {\omega}}
\newcommand{\cx}{\ensuremath {C_{X}}}

\title{The Morse Index of Redutible Solutions  \\  of the $\sw$-Equations}
\author{ Celso M. Doria \\ UFSC - Depto. de Matem\'atica}
%\data{02/10/2000}
%\email{cmdoria@mtm.ufsc.br}
%\keywords{connections,gauge fields,4-manifolds \\ MSC 58J05 , 58E50}
\maketitle
\begin{abstract}
The $2^{nd}$ variation formula of the Seiberg-Witten functional is obtained in order to 
estimate the Morse index of redutible solutions $(A,0)$. It is shown that their Morse index is given by the 
dimension of the largest negative eigenspace of the  operator $\triangle_{A} +\frac{k_{g}}{4}$,
 hence it  is finite.
\end{abstract}
%\maketitle

\section{\bf{Introduction}}

Let $(X,g)$ be a closed  riemannian 4-manifold and 

\begin{equation*}
Spin^{c}(X)=\{ \beta +\gamma\in H^{2}(X,\Z)\oplus H^{1}(X,\Z_{2}) \mid w_{2}(X)=
\alpha (mod\ \ 2)\}.
\end{equation*}

\vspace{05pt}

\noindent the space of $Spin^{c}$ structures on $X$.

Originally, the $\swa$-monopole equations in ~\ref{E:03} were not obtained through a variational 
principle, but they can be described as the stable solutons of the functional in ~\ref{E:SW02}, 
named $\swa$-functional. Since this functional  satisfies the Palais-Smale condition  it is allowed  
to related its critical points with the topology  of  
its configuration space $\Q$. As shown in ~\cite{CMD04}, by considering  the embedding of 
the Jacobian Torus 

$$i:T^{b_{1}(X)}=\frac{H^{1}(X,\R)}{H^{1}(X,\Z)}\hookrightarrow\Q,\quad b_{1}(X)=dim_{\R}H^{1}(X,\R) $$ 

\noindent into the configuration space , the variational formulation of the $\swa$-equations  gives us 
 an  interpretation for the topology of $\Q$; 

\begin{theorem}\label{T:01}
Let $(X,g)$ be a closed riemannian 4-manifold  with scalar curvature  $k_{g}$. Thus,

\begin{enumerate}
\item  if $k_{g}\ge 0$, then the gradient flow of the $\swa$-functional defines an homotopy equivalence
among $\Q$ and $i(T^{b_{1}(X)})$.

\vspace{05pt}

\item if $k_{g}<0$, then $\Q$ has the same homotopy type of $T^{b_{1}(X)}$.
\end{enumerate}
\end{theorem}

 Therefore, the existence of $\swa$-monopoles are not a consequence of the homotopy type 
of $\Q$. It is also known that they may exist for at most a finite number of classes $\alpha$ and, 
whenever  $X$ has simple type, they satisfy $\alpha^{2}=2\chi(X)+3\sigma(X)$
\footnote{$\chi(X)$-euler characteristic of $X$, $\sigma(X)$=signature of $X$.} 
corresponding, each one, to a almost complex structure on $X$.

 An application of theorem ~\ref{T:01} is that  the space $\Q-\{(A,0)\mid A\in\Aa\}$
has the homotopy type of $T^{b_{1}(X)}\times \C P^{\infty}$. 

I am grateful to Professor Clifford Taubes for sharing his knowledgment about the 
spectral properties of the operator  $\triangle_{A} +\frac{k_{g}}{4}$.

\section{Background}

In order to define the $\swa$ functional,  we  fix a $Spin^{c}$ 
structure on $X$ and also describe the Sobolev Space structure defined on the 
 configuration space.  For each $\alpha \in Spin^{c}(X)$, 
there is a representation 
$\rho_{\alpha}:SO_{4}\rightarrow \C l_{4}$ inducing  a pair of vector bundles 
$(\csa^{+},\la)$ over X, where

\begin{description}
\item[(i)] $\csa = P_{SO_{4}}\times_{\rho_{\alpha}} V = \csa^{+}\oplus
  \csa^{-}$.
 The bundle  $\csa^{\pm}$ are the positive and the negative complex spinors bundle (fibers are 
$Spin^{c}_{4}-modules$ isomorphic to $\C^{2}$). 

\item[(ii)] $\la=P_{SO_{4}}\times_{det(\alpha)} \C$.
 It is called the \emph{determinant line bundle} associated to  $\alpha$. ($c_{1}(\la)=\alpha$)
\end{description}

\noindent Thus, given $\alpha \in Spin^{c}(X)$ we associate a pair of bundles

$$\alpha \in Spin^{c}(X) \quad \rightsquigarrow \qquad(\la,\csa^{+})$$

\noindent From now on, we  considered  on $X$ a Riemannian metric $g$  and on $\csa$  an hermitian 
structure $h$.

Let  $P_{\alpha}$ be the  $U_{1}$-principal bundle over X obtained as the frame bundle 
of $\mathcal{L}_{\alpha}$ ($c_{1}(P_{\alpha})=\alpha$). Also, we consider the associated bundles 

$$Ad(U_{1})=P_{U_{1}}\times_{Ad}U_{1}\quad ad(\U)=P_{U_{1}}\times_{ad}\U,$$

\noindent where $Ad(U_{1})$ is  a fiber bundle with fiber $U_{1}$, and $ad(\mathfrak{u}_{1})$ is 
a vector bundle with fiber isomorphic to the Lie Algebra $\mathfrak{u}_{1}$. 

Let $\Aa$ be (formally) the  space of connections (covariant derivative) on
$\la$, $\vsa$  the space of
sections of $\pcsa$ and $\G=\Gamma(Ad(U_{1}))$  the gauge group acting on $\Aa\times\vsa$ as follows:

\begin{equation}\label{E:gauge01}
g.(A,\phi)=(A+g^{-1}dg,g^{-1}\phi).
\end{equation}

\vspace{05pt}

\noindent So, the action is free whenever $\phi\ne 0$, otherwise the pair $(A,0)$ has isotropy subgroup
isomorphic to $U_{1}$.  $\Aa$ is an affine space which vector space structure, after fixing
an origin, is isomorphic to 
the space $\Ok$ of $\ad$-valued 1-forms. Once a connection $\triangledown^{0}\in\Aa$ is fixed, 
a bijection $\Aa\leftrightarrow\Ok$ is explicited by
$\triangledown^{A}\leftrightarrow A$, where $\triangledown^{A}=\triangledown^{0}+A$.  
$\G=Map(X,U_{1})$, since $Ad(U_{1})\simeq X\times U_{1}$. The curvature of a 1-connection form 
$A\in\Ok$ is the 2-form $F_{A}=dA\in \varOmega^{2}(ad(\mathfrak{u}_{1}))$.

The configuration space is $\ca$. A topology is fixed on $\ca$ by considering the following 
Sobolev structures:

\begin{itemize}
\item $\mathcal{A}_{\alpha}= L^{1,2}(\Omega^{1}(ad(\U))))$;
\item $\varGamma(\csa^{+})$ = $L^{1,2}(\varOmega^{0}(X,\csa^{+}))$;
\item $\ca = \mathcal{A}_{\alpha}\times \varGamma(\csa^{+})$;
\item $\G= L^{2,2}(X,U_{1})=L^{2,2}(Map(X,U_{1}))$.\\ ($\G$ is  an 
$\infty$-dimensional Lie Group which Lie algebra is $\mathfrak{g}=L^{1,2}(X,\U)$). 
\end{itemize}

\subsection{Seiberg-Witten Monopole Equation}

In dimension 4, the vector bundle $\varOmega^{2}(ad(\mathfrak{u}_{1}))$ splits as

\begin{equation*}
\varOmega^{2}_{+}(ad(\mathfrak{u}_{1}))\oplus \varOmega^{2}_{-}(ad(\mathfrak{u}_{1})),
\end{equation*}

\vspace{05pt}

\noindent where (+) is the  seld-dual component and (-) the anti-self-dual.

\vspace{05pt}

 Fixed $\alpha\in Spin^{c}(X)$, the 1$^{st}$-order  $\sw_{\alpha}$-monopole equations, defined  over 
 $\ca =\mathcal{A}_{\alpha}\times \varGamma(\csa^{+})$, are

\begin{equation}\label{E:03}
\begin{cases}
D^{+}_{A}(\phi )= 0,\\ 
F^{+}_{A} = \sigma (\phi),
\end{cases}
\end{equation}

\noindent where
\begin{description}
\item[(i)] $D^{+}_{A}$ is the $Spinc^{c}$-Dirac operator defined on $\vsa$,
\item[(ii)] The quadratic form $\sigma:\vsa\rightarrow End^{0}(S^{+}_{\alpha})$ is defined as  

\begin{equation*}\label{E:QF}
\sigma (\phi)=\phi\otimes\phi^{*} - \frac{\mid\phi\mid^{2}}{2}.I
\end{equation*}
\end{description}

\vspace{05pt}

\noindent performs the coupling of the \emph{ASD}-equation with
the $Dirac^{c}$ operator. 

\noindent A  solution of equations ~\ref{E:03} is named as $\sw_{\alpha}$-monopole. It can 
be described as  the space $\mathcal{F}^{-1}(0)$, where  
$\mathcal{F}_{\alpha} :\ca \rightarrow \varOmega^{2}_{+}(X)\oplus
 \varGamma(\csa^{-})$ is the map

$$\mathcal{F}_{\alpha}(A,\phi) = (F^{+}_{A} - \sigma (\phi), D^{+}_{A}(\phi) ).$$ 

\vspace{05pt}

\noindent The $\sw_{\alpha}$-equations are $\G$-invariant and the map $\mathcal{F}$ is a Fredholm
map up to  gauge equivalence.

\begin{definition}\label{D:4}
$(A,\phi)$ is named a $\swa$-monopole if it satisfies the equation ~\ref{E:03} and  $\phi\ne 0$.  
\end{definition}

\subsection{$\sw_{\alpha}$ Lagrangean}

\begin{definition}
For each $\alpha\in Spin^{c}(X)$, the Seiberg-Witten Functional is the functional 
 $\sw_{\alpha}:\ca\rightarrow\R$ given by

 \begin{equation}\label{E:SW02}
\sw(A,\phi) = \int_{X}\{\frac{1}{4}\mid F_{A}\mid^{2} + \mid \triangledown^{A} \phi
\mid^{2} + \frac{1}{8}\mid \phi\mid^{4} + \frac{k_{g}}{4}\mid\phi\mid^{2} \}dv_{g} +\pi^{2}\alpha^{2},
\end{equation}
where $k_{g}$ = scalar curvature of $(X,g)$ and $\alpha^{2}=Q_{X}(\alpha,\alpha)$ ($Q_{X}$ is the 
intersection form of $X$)
\end{definition}

\begin{remark}.
\begin{description}
\item[(i)]\emph{The functional is well defined on $\ca$, since in dimension n=4 there is the 
Sobolev embeddings   $L^{4}(\Ok)\subset L^{1,2}(\Ok)$ and \\
$L^{4}(\varOmega^{0}(X,\csa^{+})\subset L^{1,2}(\varOmega^{0}(X,\csa^{+}))$.} 

\item[(ii)] \emph{the functional  $\sw_{\alpha}:\ca\rightarrow\R$ is gauge invariant.}

\item[(iii)] \emph{Whenever $k_{g}\ge 0$, $\sw_{\alpha}(A,\phi)\ge 
\int_{X}\mid F_{A}\mid^{2}dx+\pi^{2}\alpha^{2}$. Therefore, the  stable solutions  are 
 $(A,0)$.}
\item[(iv)] \emph{$(A,0)$ is a $\sw_{\alpha}$-monopole if and only if $A$ is  anti-self-dual connection. 
Whenever $b^{+}_{2}\ge 3$, it is known ~\cite{DK91} that such solutions do not exist for a 
dense set of metrics  on $X$.} 

\end{description}
\end{remark}

The tangent bundle $T\ca$ is  trivial because $\ca$ is contractible. The fiber over $(A,\phi)$ is  

$$T_{(A,\phi)}\ca =\Ok \oplus \vsa .$$

\vspace{05pt}

\noindent A riemannian structure on $T\ca$ is defined by the inner product 

\begin{align*}
<,>:&T_{(A,\phi)}\ca\times T_{(A,\phi)}\ca\rightarrow \R\\
&\left((\Theta+V),(\Lambda+W)\right)\to <\Theta+V,\Lambda +W>=<\Theta,\Lambda>+<V,W>.
\end{align*}

\noindent where

\begin{align*}
&<\Theta,\Lambda>=\int_{X}<\Theta(x),\Lambda(x)>dx,\\
& <V,W>=\int_{X}\{<V(x),W(x)>dx.
\end{align*}

\vspace{05pt}

\noindent The derivative of  $\sw$-functional defines a 1-form 
$d\sw\in\Omega^{1}(\ca)$. For each point $(A,\phi)$, the functional 
$d\sw_{(A,\phi)}:\Ok\oplus\vsa\rightarrow \R$ is decomposed in components as 

$$d\sw_{(A,\phi)}.(\Theta +\Lambda)=d_{1}\sw_{(A,\phi)}.\Theta + d_{2}\sw_{(A,\phi)}.\Lambda,$$

\vspace{05pt}

\noindent  where 

\begin{description}
\item[(i)] $d_{1}$
\begin{align*}
d_{1}\sw_{(A,\phi)}.\Theta &=  \lim_{t\to 0}\frac{\sw(A+t\Theta,\phi)-\sw(A,\phi)}{t}=\\ 
&=\frac{1}{2}\int_{X}Re\{<F_{A},d\Theta> + 
4<\triangledown^{A}(\phi),\Theta(\phi)>\}dx=\\
&=\frac{1}{2}Re\left(\int_{X}<d^{*}F_{A} +4\Phi^{*}(\triangledown^{A}(\phi)),\Theta>\right)dx=\\
&=\frac{1}{2}Re\left(<d^{*}F_{A} +4\Phi^{*}(\triangledown^{A}(\phi)),\Theta>\right).
\end{align*}

\vspace{05pt}

\noindent where
$\Phi:\varOmega^{1}(\mathfrak{u}_{1})\rightarrow\varOmega^{1}(\csa^{+})$  
is the linear operator $\Phi(\Theta) = \Theta(\phi)$, and 

\begin{equation*}
\Phi^{*}(\triangledown^{A}(\phi))=\frac{1}{2}d(\mid\phi\mid^{2}).
\end{equation*}

\item[(ii)] $d_{2}$
\begin{align*}
d_{2}\sw_{(A,\phi)}.V &=\lim_{t\to 0}\frac{\sw(A,\phi+tV)-\sw(A,\phi)}{t} =\\ 
&=2.
\int_{X}Re\{<\triangledown^{A}\phi ,\triangledown^{A}V> + 
<\frac{\mid\phi\mid^{2}+k_{g}}{4}\phi, V>\}dx=\\
&=2.\int_{X}Re\{<\triangle_{A}\phi + \frac{\mid\phi\mid^{2}+k_{g}}{4}\phi, V>\}dx=\\
&=2.Re\left(<\triangle_{A}\phi + \frac{\mid\phi\mid^{2}+k_{g}}{4}\phi, V>\right).
\end{align*}

\noindent where $\triangle_{A}=(\triangledown^{A})^{*}\triangledown^{A}$. 

\end{description}

Thus, the Euler-Lagrange equations of $\sw_{\alpha}$-functional  are

\begin{align}
d^{*}F_{A} + 4\Phi^{*}(\triangledown^{A}\phi)&=0\label{E:09},\\
\Delta_{A} \phi + \frac{\mid \phi \mid^{2}}{4}\phi +\frac{k_{g}}{4}\phi &= 0, \label{E:10}.
\end{align}

\vspace{05pt}

\subsection{Hodge Solutions}

It follows from the  identity

\begin{equation*}
\sw_{\alpha}(A,\phi)=\int_{X}\{\mid D_{A}\phi\mid^{2}+\mid F^{+}_{A}-\sigma(\phi)\mid^{2}\}dx,
\end{equation*}

\noindent (see in ~\cite{CMD06}) that the $\sw_{\alpha}$ monopoles are the stable 
solutions. From \cite{Wi94}, it is known that 
such $\sw$-monopoles exist only for a finite number of classes $\alpha\in Spin^{c}(X)$. 
Thus, it arise the question about a sufficient condition to guarantee their existence. 
The easiest solutions of the Euler-Lagrange equations above are the ones of type $(A,0)$, where
the curvature $F_{A}$ is  a harmonic 2-form.  Hodge theory guarantees the existence of such solution. 
The gauge isotropy subgroup of $(A,0)$ is isomorphic to $U_{1}$, justifying their nickname \emph{redutible}. 
Furthermore,

\begin{proposition}
 The space of solutions of  $d^{*}F_{A}=0$, module the $\G$-action, 
is diffeomorphic to  the Jacobian Torus

$$T^{b_{1}(X)}=\frac{H^{1}(X,\R)}{H^{1}(X,\Z)},\quad b_{1}(X)=dim_{\R}H^{1}(X,\Z).$$
  
\end{proposition}
\begin{proof}
\noindent The equation $d^{*}F_{A}=0$ 
implies that $F_{A}$ is an harmonic 2-form and, by Hodge theory, it is the only one. 
Let $A$ and $B$ be  solutions and consider $B=A+b$; so, 

$$d^{*}F_{B}+d^{*}F_{A} + d^{*}db=0\quad\Rightarrow db=0.$$

\vspace{05pt}

\noindent By fixing the origin at $A$, we associate  
$B\rightsquigarrow b\in H^{1}(X,\R)$; note that $F_{B}=F_{A}$. 
Let $B_{1}$ be a solution gauge equivalent to $B_{2}$, so there exists 
$g\in\G$ such that $B=A+g^{-1}dg$ and 
$F_{B_{1}}=F_{B_{2}}$. However, the 1-form $g^{-1}dg\in H^{1}(X,\Z)$. 
Consequently, if $B_{1}=A+b_{1}$ 
and $B_{2}=A+b_{2}$, then $[b_{2}]=[b_{1}]$ in 
$\frac{H^{1}(X,\R)}{H^{1}(X,\Z)}$. 
\end{proof}

\section{2nd - Variation Formula }

The $2^{nd} $ - variation formula is the  symmetric bilinear form 

$$H^{\sw}_{(A,\phi)}:T_{(A,\phi)}\ca\times T_{(A,\phi)}\ca\rightarrow \R.$$

\vspace{05pt}
 
\noindent obtained by computing the hessian of $\sw$. A neat exposition may achieved by
 considering the operator $\h$  as a linear functional 
$\h:T_{(A,\phi)}\ca\otimes T_{(A,\phi)}\ca\rightarrow\R$. 
By taking  $\mathcal{U}=\Ok$ and $\mathcal{V}=\vsa$,
we have the functional
 
$$\h:(\mathcal{U}\otimes\mathcal{U})\bigoplus (\mathcal{U}\otimes\mathcal{V})\bigoplus 
(\mathcal{V}\otimes\mathcal{U})\bigoplus (\mathcal{V}\otimes\mathcal{V})\rightarrow\R$$

\vspace{05pt}

\noindent decomposing into 

\begin{align*}
\h((\Theta,V),(\Lambda,W))&=\h(\Theta\otimes \Lambda + 
\Theta\otimes W + V\otimes\Lambda +V\otimes W)=\\
&=d_{11}(\Theta\otimes\Lambda) + d_{12}(\Theta\otimes W) + d_{21}(V\otimes\Lambda) + d_{22}(V\otimes W).
\end{align*}

\vspace{05pt}

\noindent Thus, $\h=d_{11}+ d_{12}+ d_{21}+ d_{22}$, where

\vspace{05pt}

\begin{enumerate}

\item $d_{11}:\Ok\otimes\Ok\rightarrow\R$ 

\begin{align*}
d_{11}\sw_{(A,\phi)}(\Theta,\Lambda)&=\lim_{t\rightarrow 0}\frac{1}{t}
\{d_{1}\sw_{(A+t\Lambda,\phi)}.\Theta - d_{1}\sw_{(A,\phi)}.\Theta \}=\\
&=\frac{1}{2}\int_{X}Re\{<d\Theta,d\Lambda> + 4<\Theta(\phi),\Lambda(\phi)>\}dx=\\
&=Re\{<d\Theta,d\Lambda> + 4<\Theta(\phi),\Lambda(\phi)>\}=\\
&=\frac{1}{2}Re\{<\Theta,(d^{*}d + 4\Phi^{*}\Phi)(\Lambda)>\}.
\end{align*}
%\end{align*}

\vspace{10pt}

\item $d_{12}:\Ok\otimes\vsa\rightarrow\R$

\begin{align*}
d_{12}\sw_{(A,\phi)}(\Theta,W)&=\lim_{t\rightarrow 0}\frac{1}{t}
\{d_{1}\sw_{(A,\phi + tW)}.\Theta - d_{1}\sw_{(A,\phi)}.\Theta \}=\\
&=2\int_{X}Re\{<\triangledown^{A}\phi,\Theta(W)> + <\triangledown^{A}W, \Theta(\phi)>\}dx\\
&=2.Re\{<\triangledown^{A}\phi,\Theta(W)> + <\triangledown^{A}W , \Phi(\Theta)>\}=\\
&=2.Re\{<\triangledown^{A}\phi,\Theta(W)> + <\Theta,\Phi^{*}\triangledown^{A}W>\}
\end{align*}

\vspace{05pt}

\noindent Introducing the operator $\mathcal{W}:\vsa\rightarrow\Ok$, as 
$\mathcal{W}(\Theta)=\Theta(W)$,

\begin{align*}
d_{12}\sw_{(A,\phi)}(\Theta,W)=2.Re\{<\Theta,P(W)>\}
\end{align*}

\vspace{05pt}

\noindent where $P:\vsa\rightarrow\Ok$ is $P(W)=\mathcal{W}^{*}(\triangledown^{A}\phi) + 
\Phi^{*}(\triangledown^{A}W)$.

\vspace{05pt}

\item $d_{21}:\vsa\otimes\Ok\rightarrow\R$

\begin{align*}
d_{21}\sw_{(A,\phi)}(V,\Lambda)&=\lim_{t\rightarrow 0}\frac{1}{t}
\{d_{2}\sw_{(A+t\Lambda,\phi)}.V - d_{2}\sw_{(A,\phi)}.V \}=\\
&= 2.\int_{X}Re\{<\triangledown^{A}\phi,\Lambda(V)> + <\Lambda(\phi),\triangledown^{A}V>\}dx\\
&=2.Re\{<\triangledown^{A}\phi, \Lambda(V)> + <\Phi(\Lambda),\triangledown^{A}V>\}\\
&=2.Re\{<\triangledown^{A}\phi , \Lambda(V)> + <V,(\triangledown^{A})^{*}\Phi(\Lambda)>\}
\end{align*}

\noindent Analogously, by considering the operator $Q:\Ok\rightarrow\vsa$, \\
$Q(\Lambda)=\Lambda^{*}\triangledown^{A}\phi + (\triangledown^{A})^{*}\Phi(\Lambda)$,

$$d_{21}\sw_{(A,\phi)}(V,\Lambda)=2.Re\{<Q(\Lambda),V>\}.$$
\vspace{05pt}

\noindent It is straight forward that $Q=P^{*}$.

\vspace{05pt}

\item $d_{22}:\vsa\otimes\vsa\rightarrow\R$
\begin{align*}
d_{22}\sw_{(A,\phi)}(V,W)&=\lim_{t\rightarrow 0}\frac{1}{t}
\{d_{2}\sw_{(A,\phi +tW)}.V - d_{2}\sw_{(A,\phi)}.V \}=\\
&=2\int_{X}Re\{<\triangledown^{A}W,\triangledown^{A}V> + \frac{\mid\phi\mid^{2}}{4}<W,V> +\\
&+\frac{1}{2}Re(<\phi,W>)<\phi,V> + \frac{k_{g}}{4}<W,V>\}dx=\\
=&2.Re\{<V,(\triangledown^{A})^{*}\triangledown^{A}W  + \frac{k_{g} + \mid\phi\mid^{2}}{4}W
+ \frac{1}{2}<\phi,W>\phi>\}
\end{align*}

\end{enumerate}

\vspace{05pt}

From the computation above,  there exists a linear operator
$\mathcal{H}_{(A,\phi)}:T_{(A,\phi)}\ca\rightarrow T_{(A,\phi)}\ca$, $\mathcal{H}=\mathcal{H}_{(A,\phi)}$,
 given by

$$\h((\Theta,V),(\Lambda,W))=<(\Theta,V),\mathcal{H}(\Lambda,W)>,$$ 

\vspace{05pt}

\noindent which matrix representation is 

\begin{equation}\label{E:95}
 \mathcal{H} = \left(
 \begin{matrix}
 d^{*}d + 4\Phi^{*}\Phi & P  \\
 Q & 
(\triangledown^{A})^{*}\triangledown^{A}  +  \frac{1}{2}<\phi,.>\phi +
 \frac{k_{g} + \mid\phi\mid^{2}}{4}I
 \end{matrix}
 \right). 
 \end{equation}

\vspace{05pt}

\noindent Therefore, the  induced  quadratic form  
$\hh:\Ok\oplus\vsa\rightarrow\R$ is 

\begin{align}\label{E:he}
\h(\Theta,V)=&\mid d\Theta\mid^{2} +\  \mid\Phi(\Theta)\mid^{2} +\
2.<\triangledown^{A}\phi,\Theta(V)> +\ 2 <\Phi(\Theta),\triangledown^{A}V> +\\
& + \mid\triangledown^{A}V\mid^{2} +\
\frac{k_{g}+\mid\phi\mid^{2}}{4}.\mid V\mid^{2} +\ \frac{1}{2}<\phi,V>^{2}.
\end{align} 

\vspace{05pt}
 \noindent Hence, $\h:T_{(A,\phi)}\ca\otimes T_{(A,\phi)}\ca\rightarrow\R$ is a bounded operator.

\vspace{05pt}

\subsection{Morse Index of Redutible Solutions}

\noindent From  Hodge theory, we have the complex

\begin{equation}
\Omega^{0}(\ad)\overset{d}{\longrightarrow} \Omega^{1}(\ad)
\overset{d}{\longrightarrow} \Omega^{2}(\ad)\overset{d}{\longrightarrow}\dots\Omega^{4}(\ad).
\end{equation}

\vspace{05pt}
\noindent Since $\Ok=d(\Omega^{0})\oplus ker(d^{*}(\Omega^{1}))$, and $d(\Omega^{0})$ is 
the tangent space to the orbits,
 the tangent space of $\Q$ at $(A,\phi)$ can be decomposed into 
$\mathfrak{W}_{1}\oplus\mathfrak{W}_{2}$, where

\begin{align*}
\mathfrak{W}_{1}&=\{\Theta\in\Ok;\  d^{*}\Theta=0\},\\
\mathfrak{W}_{2}&=\{W\in\vsa ;\ <W,A(\phi)>=0,\ \forall A\in\Omega^{0}(\ad) \}
\end{align*} 

\vspace{05pt}

\noindent Furthermore, $\mathfrak{W}_{1}=d^{*}(\Omega^{2}(\ad))\oplus\mathcal{H}_{1}$, 
where $\mathcal{H}_{1}$ is space of harmonic forms and  is also the tangent space to the Jacobian torus
$T^{b_{1}(X)}$ at $(A,0)$.

The quadratic form in ~\ref{E:he}, when evaluated at a redutible solution $(A,0)$,  is given by

\begin{equation}\label{E:87}
\mathcal{H}(\Theta,V)=\mid\mid d\Theta\mid\mid^{2}_{L^{2}}+ \mid\mid \triangledown^{A}V\mid\mid^{2}_{L^{2}} + 
\int_{X}\frac{k_{g}}{4}\mid V\mid^{2}dx.
\end{equation}

\vspace{05pt}

\noindent Indeed, at $(A,0)$, it follows from expression ~\ref{E:he}

\begin{equation}\label{E:95}
 \mathcal{H} = \left(
 \begin{matrix}
 d^{*}d  & 0  \\
 0 & 
\triangle_{A}  + \frac{k_{g} }{4}
 \end{matrix}
 \right),\qquad \h(\Theta,V)=<(\Theta,V),\mathcal{H}(\Theta,V)>. 
 \end{equation}

\vspace{05pt}

\noindent The  Morse index of $(A,0)$ is equal to the dimension of  the largest 
negative eigenspace  of the operator $L_{A}=\triangle_{A}+\frac{k_{g} }{4}:\varGamma(S_{\alpha})
\rightarrow \varGamma(S_{\alpha})$.  Let $\mathcal{V}_{\lambda}\subset T_{(A,0)}\Q$ be the eigenspace associated to the 
eigenvalue $\lambda$. Since $L_{A}$ is an elliptic operator, $\mathcal{V}_{0}$ has finite dimension;

\begin{proposition}
Let $(A,0)$ be a redutible solution. So,
\begin{enumerate}
\item If $k_{g}>0$, then $\mathcal{V}_{0}=T_{(A,0)}T^{b_{1}(X)}$ and so $\swa:\Q\rightarrow\R$ is a 
Morse-Bott function at $(A,0)$.

\item If $k_{g}=0$, then $\mathcal{V}_{0}=T_{(A,0)}T^{b_{1}(X)}\oplus \{V\in \vsa\mid \triangledown^{A}V=0\}$.
\end{enumerate}
\end{proposition}
\begin{proof}
It is straightforward from equation ~\ref{E:87}.
\end{proof}

\begin{proposition}
The Morse index of a  redutible solution $(A,0)$ is finite. 
\end{proposition}
\begin{proof}
It follows from the spectral theory applied to the elliptic, self-adjoint operator 
$L_{A}$ (~\cite{CL81}, ~\cite{WV84}) that the spectrum of $L_{A}$
\begin{description}
\item[(i)] is discrete, 
\item[(ii)] each eigenvalue has finite multiplicity, 
\item[(iii)] there are no points of accumulation,
\item[(iv)] there are but a finite number of eigenvalues below any given number.
\end{description}

\noindent Therefore, the spectrum of $L_{A}$ is bounded from below.

\end{proof}

\vspace{20pt}

\noindent\emph{Universidade Federal de Santa Catarina \\
    Campus Universitario , Trindade\\
               Florianopolis - SC , Brasil\\
               CEP: 88.040-900}
\vspace{10pt}

\end{document}